\documentstyle[12pt,aps,pra,amsfonts,epsf]{revtex}

\def\upcite#1{\mbox{$\!^{\cite{#1}}$}}

\newcommand{\bra}[1]{\mbox{$\langle{#1}|$}}
\newcommand{\ket}[1]{\mbox{$|{#1}\rangle$}}
\newcommand{\diracsp}[2]{\mbox{$\langle{#1}|{#2}\rangle$}}
\def\I{{\rm i}}

\def\e{{\rm e}}

\begin{document}
\draft
\title{
Quantum CPU and Quantum Simulating
\thanks{Supported by the National Natural Science Foundation of China under Grant No. 69773052 and the Fellowship of China Academy of Sciences}}

\author{An Min WANG$^{1,2,3}$}
\address{CCAST(World Laboratory) P.O.Box 8730, Beijing 100080$^1$\\
and Laboratory of Quantum Communication and Quantum Computing\\
University of Science and Technology of China$^2$\\
Department of Modern Physics, University of Science and Technology of China\\
P.O. Box 4, Hefei 230027, People's Republic of China$^3$}


\maketitle

\vspace{0.2in}

\begin{abstract}
{\em Making use of an universal quantum network or QCPU proposed by me\upcite{My1} some special quantum networks for simulating some quantum systems are given out. Specially, it is obtained that the quantum network for the time evolution operator which can simulate, in general, Schr\"odinger equation}. 
\end{abstract}
\medskip
\pacs{PACS: 03.67.Lx, 03.65-a, 03.65.Bz}

\vskip 0.1in


Simulating quantum systems has such a meaning that using a specially designed quantum system, for example quantum computer, which is called a simulated system, to simulate another so-called physical quantum system. In 1982, Richard Feynman first proposed that a quantum system would be more efficiently simulated by a computer based on the principles of quantum mechanics rather than by one based on the principles of classical mechanics.\upcite{Feynman} This is because that the size of the Hilbert space grows exponentially with increase of the number of particles. A full quantum simulating demands the exponential resources on a classical computer so that it is in general intractable. Since the discovery by Shor of a quantum algorithm for factoring in polynomial time,\upcite{Shor} there has been tremendous activity in the field of quantum computation including quantum simulating. For example, Lloyd has shown how a quantum computer is in fact an efficient quantum simulator.\upcite{Lloyd} In addition, some general ideas and schemes to several special quantum systems were proposed and discussed.\upcite{WZBALSM} At present, quantum simulating mainly performs a simulation of the dynamics. 

Obviously, it is very important how to design a quantum network for simulating Schr\"odinger equation. However, it is a difficult task by using of the known method given by Barenco {\it et.al}.\upcite{Barenco} It seems to me that the elementary difficulties are original from the facts that Hamiltonian is a summation operator of the kinetic energy and the potential energy, and the time evolution operator is a multiple product operator of a series of successive the small enough time evolution operators within a given precision. Moreover, the small enough time evolution operator is only approximately unitary in practical. Therefore one needs an alternative method which can give such a quantum network that it is easily able to construct a whole network for the summation and product of the simultaneous and/or successive transformations. In my paper,\upcite{My1} I proposed an universal quantum network for quantum computing $U$, which can be called quantum CPU and be defined as:
\begin{equation}
Q(U)=\prod_{m,n=0}^{2^k-1}\exp\{(U_{mn}\ket{m}\bra{n}\otimes I_A)\cdot C_A^\dagger\},\label{UQN}
\end{equation}
where $I_R$ and $I_A$ are the identity matrices in the register space and the auxiliary qubit space respectively, and $C_A^\dagger=I_R\otimes c_A^\dagger=I_R\otimes \ket{1}{}_A{}_A\bra{0}$. 
If the graphics rules for the factor with form $\exp\{(U_{mn}\ket{m}\bra{n})\cdot C_A^\dagger\}$ are given out, the picture of quantum network $Q(U)$ can be drawn easily. It is the most important that $Q(U)$ has two very useful properties
\begin{equation}
Q(U_1+U_2+\cdots+U_r)=Q(U_1)Q(U_2)\cdots Q(U_r),
\end{equation}
\begin{equation}
Q(U_1U_2\cdots U_r)= I_R\otimes I_A+ C_A^\dagger\left(\prod_{j=1}^{r} C_AQ(U_j)\right) C_AC_A^\dagger,\label{CQC}
\end{equation} 
where $C_A$ is the so-called ``{\it Connector}" defined as  $
C_A=I_R\otimes c_A=I_R\otimes \ket{0}{}_A{}_A\bra{1}$. It is used to the preparing transformed state so that this prepared state can be used in the successive transformation. 
Furthermore, note that there are the relations $c_A^2=c_A^{\dagger 2}=0; c_A c_A^\dagger+c_A^\dagger c_A=I_A$, 
thus $c_A$ and $c_A^\dagger$ can be thought of as the fermionic annihilate and create operators respectively in the auxiliary qubit. In order to give out the realization of QCPU for the product of transformation in a form of full multiplication, eq.(\ref{CQC}) can be rewritten as
\begin{equation}
\bar{Q}(U_1U_2\cdots U_r)=(I_R)_{input}\otimes \left[C_A^\dagger\left(\prod_{j=1}^{r} C_AQ(U_j)\right) C_AC_A^\dagger\right]_{out}
\end{equation}
while the initial state is now prepared as $(\ket{\Psi(t)})_{input}\otimes (\ket{\Psi(t)}\otimes \ket{0}_A)_{out}$.
Therefore, it seems to me this new construction of the universal quantum network is scalable easily. In terms of it, it is easy to design, assemble and scale the whole quantum networks to simulate some quantum systems such as free particle, harmonic oscillator, one particle in a constant field. In special, it is obtained in this letter that the quantum network for the time evolution operator that can simulate, in general, Schr\"odinger equation.


First, a basic skill to simulate quantum systems is to discretize the wavefunction which describes the quantum state in a finite space as the following:
\begin{equation}
\psi(x=a,t)\longrightarrow \psi(x_m,t)\quad \left(m=\left[\frac{a}{L/N}\right]\right).
\end{equation}
When the quantum system is limited within a box, it ought to impose the periodic boundary conditions, that is
\begin{equation}
\psi(x_{m+N},t)=\psi(x_m,t)\label{PBC}.   
\end{equation}
Therefore, the wavefunction in time $t$ can be written as a vector
\begin{equation}
\ket{\Psi(t)}=\sum_{m=0}^{N-1} \psi(x_m,t)\ket{x_m},
\end{equation}
where $\ket{x_m}$ form a set of basis with the properties of orthogonality and completeness ($\diracsp{x_m}{x_n}=\delta_{mn}$, $\sum_{m=0}^{N-1}\ket{x_m}\bra{x_m}=1$) in $N$-dimensional Hilbert space. For two particles, it is extended as 
\begin{equation}
\ket{\Psi(t)}=\sum_{m_1=0}^{N_1-1}\sum_{m_2=0}^{N_2-1} \psi(x_{m_1},x_{m_2},t)\ket{x_{m_1}x_{m_2}}.
\end{equation}
Of course, the extension to high dimensional is similar. In fact, the above equation also can describe two dimensional case. (Usually taking $N_1=N_2$,  this means that the particle moves in a square box).  
A quantum register with $k$ qubits can express a state in Hilbert space at most with  $N=2^k$ dimensional. For two particles in one dimensional, its Hilbert space should be $N_1\times N_2=2^{k_1+k_2}$. Thus, in a classical computer, we need exponentially increasing bits to deal with this quantum state. But in a quantum computer, we see that $k_1+k_2$ qubits, or two quantum registers with $k_1$ and $k_2$ qubits respectively, are suitable to this task. In fact, this is just one of reasons why simulating a quantum system can be more rapidly done by use of a quantum computer than by use of a classical computer. Generally speaking, for 3-dimensional and $n$ particles, we can use $3n$ quantum registers to store the discretizing wavefunction. For the system of identical particles the initial state of the quantum computer has to be chosen symmetrically or anti-symmetrically. For quantum field theory, its discretized method can be similar to one in lattice gauge theory. 

It is worth emphasing how to realize the fundamental operators such as coordinates and momentum is a key point. In coordinate representation, coordinates are directly written as a diagonal matrix whose diagonal elements are $x_m$. But the momentum has a little complication because it is a derivative action as the following
\begin{equation}
\hat{p}\psi(x,t)=-\I\frac{\partial \psi(x,t)}{\partial x}=-\I\frac{\psi(x+\triangle x,t)-\psi(x)}{\triangle x}.
\end{equation}
So the momentum operator can be defined by:
\begin{equation}
\hat{p}=-\I\;\frac{1}{2}\left(\frac{N}{L}\right)\sum_{m=0}^{N-1}(\ket{x_m}\bra{x_{m+1}}-\ket{x_m}\bra{x_{m-1}}),
\end{equation}
where, in order to make the momentum operator is Hermian,  an average of the left and right derivative has been taken and the periodic boundary conditions (\ref{PBC}) has been used. Thus 
\begin{equation}
\hat{p}\ket{\Psi(t)}=\sum_{m=0}^{N-1}\left[-\I\;\frac{1}{2}\frac{\psi(x_{m+1},t)-\psi(x_{m-1},t)}{(L/N)}\right]\ket{x_m}.
\end{equation}

Further, the kinetic energy operator (natural unit system $\hbar=c=1$) is obtained:
\begin{equation}
\hat{T}=-\frac{1}{8\mu}\left(\frac{N}{L}\right)^2\left[\sum_{m=0}^{N-1}(\ket{x_m}\bra{x_{m+2}}+\ket{x_m}\bra{x_{m-2}})-2I_N\right].\label{Tenergy}
\end{equation}

It is clear that the potential operator is a diagonal transformation if one only considers the local potential $U(x)$ or the external field $V_e(x)$, that is
\begin{eqnarray}
\hat{U}&=&\sum_{m=0}^{N-1}U(x_m)\ket{x_m}\bra{x_m},\\
\hat{V}_e&=&\sum_{m=0}^{N-1}V_e(x_m)\ket{x_m}\bra{x_m},
\end{eqnarray}        
while the two-body local interaction $U(x_1,x_2)$ can be written as
\begin{equation}
\hat{U}=\sum_{m_1}^{N_1-1}\sum_{m_2}^{N_2-1}U(x_{m_1},x_{m_2})\ket{x_{m_1},x_{m_2}}\bra{x_{m_1},x_{m_2}}.
\end{equation}

It is easy to extend to high dimensional and many particles formally. For example, in two particles case, the fundamental momentum operators are $\hat{p}_1=\hat{p}^{(1)}\otimes I_2, \hat{p}_2=I_1\otimes \hat{p}^{(2)}$. Then, the kinetic energy and the potential energy operators can be constructed in similar way.

Choosing a short time step $\triangle t$, the time evolution operator $\e^{-\I Ht}$ reads
\begin{equation}
\Omega(T)=(1+iH\triangle t)^{T/\triangle t}.
\end{equation}
where Hamiltonian $H=T+V$ can be given out as above. 



Now, we have the ability to simulate some the quantum systems. First, let's consider a simple example -- a free particle in one dimensional. 
Its Hamiltonian is $H=p^2/2\mu$. It is convenient to simulate it in the momentum representation. So we can use two quantum networks to simulate its dynamics. One is the quantum network $Q(F)$ for the discreting Fourier transformation, the  
other is the quantum network $Q(\e^{\I Ht})$ for the diagonal transformation. Then, we use the connector $C_A$ to combine them in turn
\begin{equation}
\bar{Q}_{\rm FP}(\e^{\I Ht}F)=(I_R)_{input}\otimes \left(C_A^\dagger C_AQ(\e^{\I Ht})C_AQ(F)C_AC_A^\dagger\right)_{out}.
\end{equation}
 
It is easy to realize the quantum network for a diagonal unitary transformation by the QCPU. When the diagonal matrix is $\e^{\I p^2t/2\mu}$, it follows that 
\begin{equation}
Q(\e^{\I p^2t/2\mu})=\prod_{m=0}^{2^k-1}\exp\{(\e^{\I p^2_mt/2\mu} \ket{x_m}\bra{x_m}\otimes I_A)\cdot C^\dagger\}.
\end{equation}
While in my paper\upcite{My3}, the quantum network for quantum Fourier transformation on a $k-$qubit register has been obtained:
\begin{equation}
Q(F)=\prod_{n=0}^{2^k-1}Q[B(n)HM_{0n}]= \prod_{m=0}^{2^k-1}\prod_{n=0}^{2^k-1}\exp\{[(B(n)H)_{m0}\ket{m}\bra{n}\otimes I_A] \cdot C^\dagger\},
\end{equation}
where $B(n)H=\prod_{\otimes,j=0}^{2^k-1}B_j[2^j\pi n/(2^k-1)]]H_j$ and  $M_{0n}=\ket{0}\bra{x_n}$. 

Now, a free particle in one dimensional can be simulated. But, the above method is perhaps not the simplest. It can be seen in the following, by virtue of the kinetic energy quantum network proposed by this letter the free particle easier can be simulate more easily. 


Secondly, for a harmonic oscillator in one dimensional, its Hamiltonian is $H=\sum_n\omega(n+1/2)\ket{n}\bra{n}$. Because it is diagonal, its realization of QCPU is easy to get:
\begin{equation}
Q_{HO}=\prod_{m=0}^{2^k-1}\exp\{\I \omega(m+1/2)t(\ket{x_m}\bra{x_m}\otimes I_A)\cdot C_A\}.
\end{equation} 
Its action can simulate dynamics of a harmonic oscillator. Obviously, simulating a harmonic oscillator is much simpler than a free particle since an appropriate representation leads in the diagonal Hamiltonian is chosen. This implies that 
the picture theory in quantum mechanics can simplify the problem. For example, one would like to simulate one particle in one dimensional under a constant field $u$. Its Hamiltonian is $H=p^2/2\mu+u$. To do this, the interaction picture can be used and then  $V_I=\e^{iH_0t}u\e^{-iH_0t}=u$ and $\psi_I(t)=\e^{iut}\psi$. Just like done for the free particle, it can be simulated easily.


The last, in order to simulate a general quantum system, it is necessary to know how to simulate Schr\"dinger equation. For simplicity, let's first consider a local potential or an external field. The quantum networks for the kinetic energy and the potential energy can be constructed as:
\begin{eqnarray}
Q(T)&=&\prod_{m=1}^{2^k-1}\exp\left\{-\frac{1}{8\mu}\left(\frac{L}{N}\right)^2(\ket{x_m}\bra{x_m}E(m,m+2)\otimes I_A)\cdot C_A^\dagger\right\}\\ \nonumber 
& &\exp\left\{-\frac{1}{8\mu}\left(\frac{L}{N}\right)^2(\ket{x_m}\bra{x_m}E(m,m-2)\otimes I_A)\cdot C_A^\dagger\right\}\cdot Q\left(\frac{1}{4\mu}\frac{L^2}{N^2} I_R\right),
\end{eqnarray}
\begin{equation}
Q(V)=\prod_{m=0}^{2^k-1}\exp\{(V(x_m)\ket{x_m}\bra{x_m}\otimes I_A)\cdot C_A^\dagger\},
\end{equation} 
where $E(m,n)$ is a general exchange transformation defined in\upcite{My1}.

Only consider a small time step, the realization of QCPU for the time evolution operator reads
\begin{equation}
Q(U(\triangle t))=Q(I_R)Q(\I\triangle t T)Q(\I\triangle t V).
\end{equation}
For a finite time, take the product of all the time evolution operators at small enough time steps and obtain finally the whole quantum network for the time evolution operator: 
\begin{equation}
\bar{Q}(\e^{\I Ht})= \left(I_R\right)_{input}\otimes\left[C_A^\dagger\left(\prod_{i=1}^{[t/\triangle t]} C_A Q(\Omega(\triangle t))\right)C_AC_A^\dagger\right]_{out}.
\end{equation}
By using of it, Schr\"odinger equation can be simulated in general. 

For two particles, the extension of method is direct, but it is not efficient enough in the use of computing resources if one does it directly. Although the quantum network for the kinetic energy operator is obtained by the same method, but the quantum network for the two body potential needs all the $2^k\times 2^k$ basic elements. Thus, a better method is first to reduce Schr\"odinger equation to the mass center and the relative coordinate system. In the mass center system, we need to simulate a free practice, and in the relative coordinate system, we need to simulate Schr\"odinger equation in a single-body potential. They are both able to be done just stated above. Of course, the efficiency problem here is said with respect to the comparison among the different quantum algorithms. Because, the quantum network is built in quantum parallelism, that is, it acts on all the states at the same time. Therefore, with respect to classical computing, it must be efficient. In this sense, the above method to simulate Schr\"odinger equation can be extended to higher dimensional and more particles.    
      
In conclusion, our QCPU and its realizations can be used to various problems of quantum simulating. Likewise, it can be used to various problems of quantum algorithms. However, it is important to find the special quantum network which can reach at higher or the highest efficiency. In order to do this, we need to use the fundamental laws of physics, specially the principles and features of quantum mechanics, for example coordinate system choice, representation transformation, picture scheme and quantum measurement theory if we have thought a quantum computing task to be a physical process. Moreover, to simplify the realization of QCPU for all computing steps and find the optimized decomposition, we have to use the symmetry properties of every step $U^i$ as possible. In this letter, such some examples have been given out. Some other examples in quantum algorithm also were obtained in my paper.\upcite{My3} This research is on progressing.  
 
\bigskip
  
I would like to thank Artur Ekert for his great help and for his hosting my visit to center of quantum computing in Oxford University.

\end{document}